\documentclass[12pt,preprint]{aastex}
\usepackage{natbib}
\shorttitle{}
\shortauthors{Paulo C. Cortes}

\shorttitle{Interferometric mapping of magnetic fields}
\shortauthors{P. C. Cortes}

\begin{document}

\title{Interferometric Mapping of Magnetic Fields: G30.79 FIR 10}
\author{P. Cortes}
\affil{Astronomy Department, University of Illinois at
    Urbana-Champaign, IL 61801}
\author{R. M. Crutcher}
\affil{Astronomy Department, University of Illinois at
    Urbana-Champaign, IL 61801}

\begin{abstract}
We present polarization maps of G30.79 FIR 10 
(in W43) from thermal dust emission
at 1.3 mm  and from CO J=$2 \rightarrow 1$ line emission.
The observations were obtained using the Berkeley-Illinois-Maryland
Association array in the period 2002-2004.
The G30.79 FIR 10 region shows
an ordered polarization pattern in dust emission, 
which suggests an hourglass
shape for the magnetic field. Only marginal detections
for line polarization were made from this region. Application
of the Chandrashkar-Fermi method yielded $B_{pos} \approx 1.7$ mG
and a statistically corrected mass to magnetic flux ratio
$\lambda_{C}  \approx 0.9$, or essentially critical.
\end{abstract}

\keywords{ISM: magnetic fields — ISM:polarization — stars: formation}

\section{Introduction}

The star formation process
involves a number of physical
parameters, of which the magnetic field is the least observed.
Magnetic field observations are divided into measurements of the Zeeman
effect (in order to obtain the magnetic field strength in the line
of sight), and linear polarization observations of dust and spectral-line emission.
Polarization of dust emission is believed to be perpendicular to the magnetic field
under most conditions \citep{Lazarian2003}; hence, polarization of dust emission
has been used as a major probe for the magnetic field geometry.
In order to efficiently map the polarization of dust emission and infer
information about the magnetic field morphology, high resolution
observations are required. The BIMA millimeter interferometer
has been used previously to obtain high-resolution polarization maps
in several star forming cores \citep{Rao1998,Girart1999b,Lai2001,
Lai2002,Lai2003}.
These results show fairly uniform polarization morphologies
over the main continuum sources, suggesting that magnetic fields
are strong, and therefore cannot be
ignored by star formation theory. However, the number of star formation
regions with maps of magnetic fields remains small, and every new
result is significant.

Spectral line linear polarization has been suggested to arise
from molecular clouds under
anisotropy conditions \citep{Goldreich1981}. The prediction suggests
that a few percent of
linearly polarized radiation should be detected from molecular clouds and
circumstellar envelopes in the presence of a magnetic field. It is also
predicted that the molecular line polarization will be either parallel
or perpendicular to the magnetic field, depending on the angles between
the line of sight, the magnetic field, and the anisotropic excitation
direction \citep{Goldreich1982}. This
process is known as the Goldreich - Kylafis effect.

In order to use these techniques,
we mapped the massive star forming region G30.79 FIR 10 with the BIMA array.
We measured continuum polarization at 1.3 mm and
CO $J=2\rightarrow 1$ line polarization obtaining high resolution
interferometric maps for both measurements.

The remainder of this paper is divided in five major sections. Section 2 reviews
information about the source, section 3 describes the observation
procedure.
Section 4 presents the results, section 5 gives the discussion,
and section 6 the conclusions and summary.

\section{Source Description}

G30.79 is a large dust continuum source located within the W 43 region, which is
an H II region-molecular cloud complex near $l=31^{\circ}, b=0^{\circ}$.
We observed G30.79 FIR 10, which is a massive and  dense component in the G30.79
complex.
\citet{Liszt1995} observed G30.79 in HCO$^{+}$ and $^{13}$CO, concluding
that the presence of several rings and shells in the dense  molecular gas
was a disturbance product of star formation. \citet{Vallee2000}
observed the dust continuum emission in this source at 760 $\mu$m with the JCMT
telescope. They found linear polarization of about 1.9\% with a position
angle (P.A.) of 160$^{\circ}$ at FIR 10.
\citet{Mooney1995} also observed this source at 1.3 mm
using the IRAM 30-meter antenna, detecting a total flux of 13.6 Jy; their
wide field map shows FIR 10 and the extended H II region in the G30.79 complex.
H$_{2}$O masers have also
been observed toward this region \citep{Cesaroni1988} that are within a half
arcsecond of the peak in the \citet{Mooney1995} map and
might be a signature of massive star formation.
No centimeter radio-continuum emission seems to be associated with
FIR 10, which suggests that the source is in an early stage of evolution.
\citet{Motte2003} mapped the W43 main complex in dust continuum emission
at 1.3 mm and 350 $\mu m$ with the IRAM 30-m telescope and the CSO, 
respectively.
They also mapped the HCO$^{+}$ $J=3\rightarrow2$ line and measured 
H$^{13}$CO$^{+}$ $J=3\rightarrow2$ towards prominent dust maxima.  
One of the maxima, W43-MM1, is the compact fragment we mapped with BIMA.
\citet{Motte2003} found v$_{lsr}= 98.8$ km s$^{-1}$, $\Delta v=5.9$ km s$^{-1}$
(from H$^{13}$CO$^{+}$), T$_{\textnormal{dust}} \sim 19$ K, M$\sim 3600$ 
M$_{\sun}$, and $n(\textnormal H_{2}) \sim 8.8 \times 10^{6}$ cm$^{-3}$.
They estimated  the virial mass to be M$_{vir} \sim 1000$ M$_{\sun}$,
suggesting  that this compact fragment should be in a state of gravitational
collapse unless there are other sources of support  than kinetic energy.
The W43 region, is therefore, an excellent region for the study of
the earliest stages of massive star formation.

\section{Observation Procedure}

We observed G30.79 FIR 10 between October 2002 to May 2004, 
mapping the continuum
emission at 1.3 mm and the CO $J=2 \rightarrow 1$ molecular line (at 230 GHz);
four tracks with  the BIMA array in C configuration were obtained.
We set  the digital correlator in mode 8
to observe both the continuum and the  CO $J=2 \rightarrow 1$ line
simultaneously. The 750 MHz lower side band was combined
with  700 MHz from the upper side band to
map the continuum emission, leaving a
50 MHz window for the CO line observation (at a resolution of 1.02 km s$^{-1}$).
Each BIMA telescope has a single receiver, and thus the two polarizations
were observed sequentially.  A quarter wave plate to select either right (R)
or left (L) circular polarization was alternately switched into the signal path
ahead of the receiver.  Switching between polarizations was sufficiently
rapid (every 11.5 seconds) to give essentially identical uv-coverage.
Cross-correlating the R and L circularly polarized signals
from the sky gave RR, LL, LR, and RL for each interferometer baseline,
from which maps in the four  Stokes parameters were produced.
The source 1743-038 was used as phase calibrator for G30.79 FIR 10.
The instrumental polarization was calibrated by observing
3C279,
and the ``leakages" solutions were calculated from this observation.
We used the same calibration procedure
described by \citet{Lai2001}.

The Stokes images I, U, Q and V were obtained by Fourier transforming
the visibility data using natural weighting.
The MIRIAD \citep{MIRIAD1995} package was used for data reduction.
G30.79 FIR 10 is close to the equator; therefore, we expect strong
sidelobes in the beam pattern. We followed \citet{Chernin1995}, who
observed NGC2071IR (also a source close to the equator)
with the BIMA array, and imaged only out to 20$^{\prime \prime}$
radius, due to the strong sidelobes.

\section{Observational Results}

\subsection{1.3 mm Continuum}

\begin{figure}
\label{1}
\figurenum{1}
\includegraphics[angle=-90,scale=0.7]{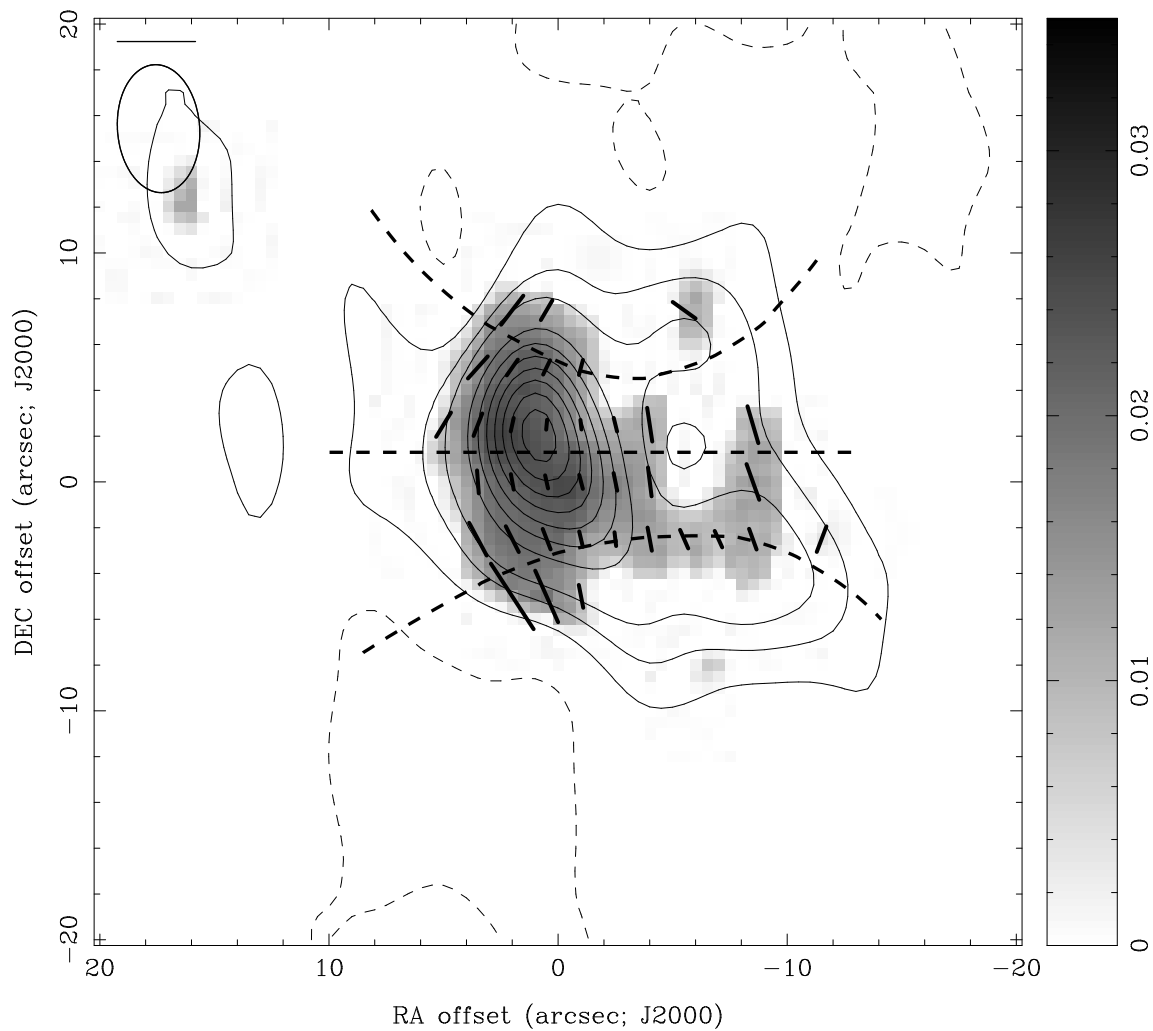}
\epsscale{0.4}
\caption{
Polarization map of G30.79 FIR 10 at 1.3 mm. The contours represent
the Stokes I emission at
-2, 2, 5, 8, 15, 20, 25, 30, 35, 40, and 45 $\times \sigma$;
 $\sigma=0.034$ Jy/beam.
The pixel gray scale shows $3\sigma$ polarized
intensity ($\sqrt{Q^{2} + U^{2}}$) for
the dust continuum emission, while the black line segments
show the fractional polarization and P.A.
The scale for the fractional polarization is given by the bar at the
left corner that represents 0.1 fractional polarization. The thick dashed
 lines
show the proposed magnetic field morphology.
}
\label{G30con}
\end{figure}

Polarized dust emission was detected and mapped toward G30.79 FIR 10
(Figure~\ref{G30con}) with a beam
size of $5.6^{\prime \prime}\times 3.6^{\prime \prime}$ and a beam
P.A. of $4.3^{\circ}$. The
Stokes I map shows a peak emission of 1.6 Jy beam$^{-1}$; 
this peak is consistent with \citet{Mooney1995}
observations at the same wavelength.
Most of the emission in Figure~\ref{G30con} comes from a compact core
 $\sim 8^{\prime \prime}$ in radius.

The polarized emission shown in Figure~\ref{G30con} has a $3\sigma$ 
level of significance,
a peak polarized intensity of 0.03 Jy beam$^{-1}$, and a mean position
angle of 4.6$^{\circ} \pm 6.1^{\circ}$.
The polarization direction is different in the northern and southern
regions;
the mean P.A. in
the north is -33.2$^{\circ} \pm 6.6^{\circ}$, and in the
south is 23.9$^{\circ} \pm 5.5^{\circ}$.
The polarization is fairly uniform in fractional polarization and
smooth in its direction change from north to south,
(see Table \ref{TableG30}).
%This
%56$^{\circ}$ difference in P.A. suggests a change in the
%magnetic field direction (assuming magnetic alignment).
\citet{Vallee2000} observed G30.79 FIR 10 using the JCMT
telescope at $760$ $\mu$m. They detected continuum polarization at two
points in FIR 10. Only one of them is in the general area of our map;
however, this point is $\sim 15^{\prime \prime}$ south of the
center of our polarization pattern. Hence, these
results cannot be compared directly with ours.

%Figure 6 shows the angle distribution histogram, 75\% of the P.A.s are
%between -30$^{\circ}$ to $40^{\circ}$ which is similar to regions
%like NGC 1333 IRAS 4 in which a similar clustering of P.A.s
%suggested a poloidal magnetic field \citep{Girart1999b}. Table 2 shows
%P.A. at each relevant pixel coordinate.

\subsection{CO $J=2 \rightarrow 1$}

\begin{figure}
\label{2}
\figurenum{2}
\includegraphics[angle=-90,scale=0.7]{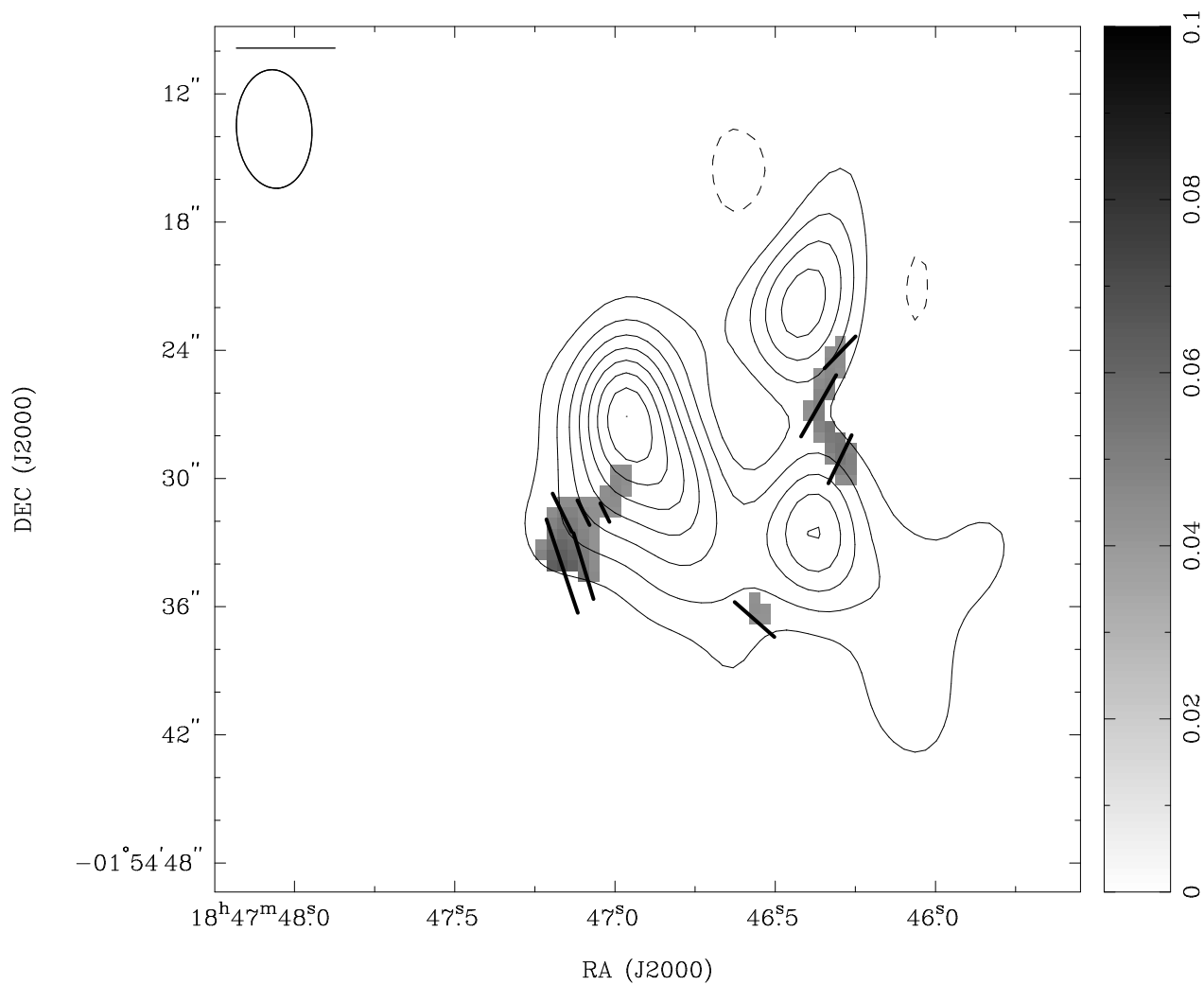}
\epsscale{0.4}
\caption{
Polarization map of G30.79 FIR 10 for the CO $J=2 \rightarrow 1$ line.
The contours represent
the Stokes I emission at -0.17, 0.17, 0.35, 0.52, 0.7, 0.88, 1.05, and 1.23
Jy/beam.  The pixel gray scale shows $3\sigma$ polarized
intensity ($\sqrt{Q^{2} + U^{2}}$) for
the CO $J=2 \rightarrow 1$ emission, while the black line segments
are the polarized line segment  map for the CO $J=2 \rightarrow 1$.
The scale for the line segments is given by the bar at the
left corner that represents 0.31 in fractional polarization.
}
\label{G30co}
\end{figure}

Polarized emission from the CO $J=2 \rightarrow 1$
was detected (Figure~\ref{G30co}). The Stokes I map shows three CO peaks
which cover a similar area as the continuum emission.
Our peak
emission is 1.4 Jy beam$^{-1}$ with a beam size of
 $5.5^{\prime \prime}\times 3.5^{\prime \prime}$ and a beam P.A. of
$4.0^{\circ}$. We only detected a level of polarized emission of
at least $3\sigma$ 
at three independant positions in the core. These three points show two
distinctive directions which appear to be orthogonal to each other.
The polarization in the southern part of the map shows a mean P.A. of
$25.5^{\circ} \pm 4.4^{\circ}$, while the north-western part
of the core shows a mean P.A. of $-31.7^{\circ} \pm 4.5^{\circ}$.
Line polarized emission appears to be parallel to dust polarized
emission over the same area in the south; the average
P.A. difference is $\sim 2^{\circ}$ which suggests that the magnetic
field is perpendicular to the polarization. 
However, in the north-western region
the average difference in P.A. is $\sim 54^{\circ}$. It is posible
that our interferometric observations are resolving out the CO emission,
which will make the fractional polarization measurements unreliable.

\section{Discussion}

We mapped polarized dust emission at 1.3 mm from G30.79 FIR 10, 
and  \citet{Vallee2000} detected polarized dust emission at 760 $\mu$m.
The emission at both wavelengths seem to be dominated by dust, which is
reinforced by the lack of centimeter radio continuum emission from
this source.  A fully developed
H II region is present in the G30.79 region, but its ionization
front has not yet reached FIR 10;
the H~II region appears to be $\sim 2$ pc away from the FIR 10 source \citep{Mooney1995}.
Water masers have also been observed, which is a signpost for high mass
star formation. Based on this evidence, we believe that G30.79 FIR 10
is in an early stage of evolution.

We inferred the source parameters from our data using an area
$\theta_{s}=20^{\prime \prime}
\times 20^{\prime \prime}$, which corresponds to the area over which we 
mapped the polarized flux of the core. 
The dust absorption cross
section per H-atom is approximated by a power law  \citep{Mezger1990}

\begin{equation}
\tau_{\lambda}=N_{\textnormal{H}} \sigma_{\lambda}^{\textnormal{H}}  \\
              =N_{\textnormal{H}} (Z/Z_{\sun}) b 
                 (7 \times 10^{-21} \lambda^{-2}),
\end{equation}

\noindent where $\lambda$ is the wavelength,
$N_{{\textnormal{H}}}=N({\textnormal{H}}) +
2N({\textnormal{H}}_{2})$
is the total hydrogen column density, $Z/Z_{\sun}$ the relative metalicity, and
$b$ is a parameter which reflects 
the variation of dust absorption cross sections
\citep{Mezger1990}.
Two values are used for $b$; $b=1.9$ reproduces estimates of
$\sigma_{400\mu m} \sim 8.3
\times 10^{-26}$ cm$^{2}$(H-atom)$^{-1}$, which represents molecular gas
of moderate density $n({\textnormal H}_{2}) < 10^{6}$ cm$^{-3}$, and $b=3.4$
relates to
dust around deeply embedded IR sources at higher densities. Apparently a value
of $b=3.4$ was used by \citet{Mooney1995} in their estimates on G30.79 FIR 10.
We used the  expressions for the cloud mass and the column density given by
\citet{Mooney1995}:

\begin{equation}
\label{cd}
(N_{\textnormal{H}}/{\textnormal{cm}}^{-2})=1.93 \times 10^{15}
\frac{(S_{\nu ,avg}/Jy)\lambda^{4}_{\mu m}}
{(\theta_{s}/arcsec)^{2}(Z/Z_{\sun})bT}
\frac{e^{x} - 1}{x}
\end{equation}
\begin{equation}
(M_{\textnormal{H}})=4.1 \times 10^{-10}
\frac{(S_{\nu ,avg}/Jy)\lambda^{4}_{\mu m} D^{2}_{kpc}}
{(Z/Z_{\sun})bT}
\frac{e^{x} - 1}{x},
\end{equation}

\noindent where $S_{\nu ,avg}$ is the averaged flux from the source, 
$\theta_{s}=
\sqrt{\theta_{s,min} \times \theta_{s,max}}$ is the 
angular source size, $x=\frac{1.44 \times 10^{4}}
{\lambda_{\mu m} T}$ is the $\frac{hc}{\lambda kT}$ factor for the Planck
function, and $D_{kpc}$ is the distance to the source in kpc.
We obtained 
$S_{\nu ,avg}$ by averaging the continuum emission over an area of
20$^{\prime \prime}
\times$ 20$^{\prime \prime}$, which gives 4.7 Jy. From their 1.3 mm single dish
observations, \citet{Mooney1995} obtained 13.6 Jy for a source
of $25.4^{\prime \prime} \times 33.2^{\prime \prime}$. Their larger result
is due both to the larger area over which they averaged and to loss of flux
in the interferometer data due to missing short uv-spacings.
We used $T=20$ K, the same dust temperature used by \citet{Mooney1995},
who  noted the uncertainty of the value, which is based on dust models.
\citet{Motte2003} also observed G30.79 FIR 10 (which they called W43-MM1); they 
fitted a gray body model obtaining a dust temperature of 19 K.
A distance of 5.5 kpc to G30.79 FIR 10 was used \citep{Motte2003}. 
From our date, we find $N_{\textnormal H}=1.3 \times 10^{24}$ cm$^{-2}$ and 
$M_{\textnormal H} = 3300$  $\textnormal{M}_{\sun}$. The mass is
in agreement with \citet{Motte2003} who estimated 
$M_{\textnormal H}=3600$  $\textnormal M_{\sun}$. However, this value is 
lower than the values calculated by \citet{Vallee2000},
 who estimated a mass
$M_{\textnormal H}=9 \times 10^{3} M_{\sun}$, and  \citet{Mooney1995},
 who calculated
$M_{\textnormal H}=1.1 \times 10^{4} M_{\sun}$. This difference
can be explained due to the larger distance assumed by the later investigators. 

Assuming magnetic alignment, our polarized dust emission
map suggests a poloidal morphology for the
magnetic field. This is seen by rotating the line segments in Figure
\ref{G30con} by 90$^{\circ}$.
This poloidal morphology hints at an hourglass shape for the magnetic field
in this core. The proposed field morphology is shown by thick dashed lines
Figure \ref{G30con}.

Using the Chandrasekhar-Fermi effect \citep{Chandrasekhar1953},
we can estimate the
magnetic field strength from the analysis of the small scale randomness
of magnetic field lines. We used the expression in \citet{Crutcher2004a} 
to estimate the magnetic field strength: 

\begin{equation}
B_{pos}=9.3\frac{\sqrt{n(\textnormal H_{2})}\Delta V}{\delta \phi}
\end{equation}

\noindent where $B_{pos}$ is the magnetic field strength in $\mu$G,
$n(\textnormal{H}_{2})$ is in cm$^{-3}$, $\Delta V$
is the FWHM velocity width in km s$^{-1}$, and $\delta \phi$ is the P.A. 
dispersion in degrees.

We used  $\Delta V=5.9$ km s$^{-1}$ from the apparently optically thin 
H$^{13}$CO$^{+}$ measured with a 29$^{\prime \prime}$ beam \citep{Motte2003}. 
Our $N_{\textnormal H}=1.3 \times 10^{24}$ cm$^{-2}$ implies 
$N$(H$_{2}$)=6.5$\times 10^{23}$ cm$^{-2}$, which gives
$n(\textnormal{H}_{2})=4.8 \times 10^{5}$ cm$^{-3}$ for a core
diameter of 16$^{\prime \prime}$ (0.43 pc).
The weighted-mean P.A. dispersion $\delta \phi=$21.9$^{\circ}$
 (after correction for 4.7$^{\circ}$ contribution from measurement
uncertainties) was calculated 
from our polarized dust observations. These values yield 
$B_{pos}=1.7$ mG. Although, this value may seem extraordinarily high,
it is similar to field strengths estimated from maser Zeeman splitting
at similar gas densities in other clouds.

Using our estimate above for the column density, we can calculate the
mass to magnetic flux ratio in term of the critical value \citep{Crutcher2004b}:

\begin{equation}
\lambda = 7.6 \times 10^{-21} \frac{N(\textnormal H_{2})}{B}
\end{equation}

\noindent where $N(\textnormal H_{2})$ is in cm$^{-2}$ 
and B in $\mu \textnormal{G}$.  We obtained 
a value of $\lambda \approx 2.8$, which is slightly supercritical.
Applying the statistical geometrical correction factor of 1/3
\citep{Crutcher2004b}, $\lambda_{C} \approx 0.9$, which is essentially
critical.

Our CO $J=2 \rightarrow 1$ line polarization detections are too scarce
to yield any significant information about
the magnetic field. However, comparing the CO polarized emission to the
dust polarized emission may help in getting information about the sources
of anisotropy in this region.
Figure~\ref{G30co} shows that at the south-eastern
side of the map  the CO polarization is parallel to the polarized
dust emission,
and in the north-western part of the map they may be roughly orthogonal.
Previous polarization observations in sources like
DR21(OH) showed CO $J=2 \rightarrow 1$ polarization perpendicular to the
polarization of dust emission over the main sources \citep{Lai2003}.
A theoretical study by
\citet{Cortes2005}  suggested that line polarization traces
the magnetic field at
lower densities than the dust, and the orientation of the line polarization
will depend on the degree of anisotropy in the region. \citet{Cortes2005}
 concluded that
the presence of hot continuum sources will preferentially produce
CO $J=2 \rightarrow 1$ polarization that will be perpendicular to the dust
polarization. In the case of G30.79 FIR10 it is difficult to draw any conclusion
without CO $J = 1\rightarrow0$ polarization data and a detailed modeling
of the source.

\section{Summary and conclusions}

We observed G30.79 FIR 10 and successfully mapped
 CO $J=2 \rightarrow 1$ line and 1.3 mm dust continuum polarized emission
with a resolution of $4^{\prime \prime}$.

G30.79 FIR 10 is not a well studied region; however, there is
evidence that points toward an early stage of development of 
high-mass star formation. We found
 a remarkably uniform pattern in our polarized dust emission map, 
which suggests
an hourglass magnetic field morphology. Using the Chandrasekhar-Fermi
method, we inferred a plane-of-the-sky magnetic field strength 1.7 mG,
which yielded a geometry-corrected mass to magnetic flux ratio of
0.9 to respect to critical. This result is similar to those found 
in many other regions of low-mass and high-mass star formation.

Polarized line emission was also
detected in this region; these results pose questions about the
sources of anisotropy, which will require more detailed modeling.

This research was partially funded by NSF grants AST 02-05810 and 02-28953.

\bibliography{biblio}

{\normalsize \begin{deluxetable}{ccc}
\tablecolumns{4}
\tablewidth{0pc}
\tablenum{1}
\tablehead{
\colhead{Offsets in arcsec} & 
\colhead{$P_{Dust}$} &
\colhead{$\phi_{Dust}$} 
}
\label{TableG30}
\tablecaption{Fractional polarization and 
Position angle for dust
polarization observations at G30.79 FIR 10. 
Data was interpolated at a tolerance of 0.5$^{\prime \prime}$
 that corresponds to approximatelly 1.11  $\times$
10$^{-2}$ pc using a distance to G30.79 of 4.6 kpc.}
\startdata
(-7.0,-9.0)  & 0.12$\pm$0.05   & 62.3$\pm$9.9 \\
(0,-6.0)     & 0.09$\pm$0.03   & 20.1$\pm$8.3   \\
(3.0,-2.0)   & 0.04$\pm$0.009  & 24.8$\pm$6.0 \\
(0,-2.0)     & 0.023$\pm$0.004 & 17.7$\pm$5.0 \\
(3.0,0)      & 0.03$\pm$0.005  & 6.6$\pm$5.3 \\
(0,0)        & 0.02$\pm$0.003  & 13.3$\pm$4.3 \\
(-2.5,0)     & 0.02$\pm$0.005  & 11.5$\pm$6.5 \\
(3.0,3.0)    & 0.03$\pm$0.004  & -21.8$\pm$4.6\\
(0,3.0)      & 0.01$\pm$0.003  & -6.6$\pm$5.7 \\
(3.0,6.0)    & 0.05$\pm$0.01   & -42.5$\pm$6.8 \\
(0,6.0)      & 0.03$\pm$0.006  & -24.3$\pm$6.7 \\
\enddata
\end{deluxetable}
}

\newpage

\end{document}